\documentclass[aps,preprint]{revtex4}
\usepackage{graphicx}% Include figure files
\usepackage{dcolumn}% Align table columns on decimal point
\usepackage{bm}% bold math
\begin{document}
%%%%%%%%%%%%%%%%%%%
%\newcommand{\kv}{\mbox{\boldmath{$k$}}}
\newcommand{\Hso}{H_{\rm so}}
\newcommand{\Hp}{H'}
\newcommand{\sigmav}{\bm{\sigma}}
\newcommand{\kv}{{\bf{k}}}
\newcommand{\qv}{{\bf{q}}}
\newcommand{\Sv}{{\bf{S}}}
\newcommand{\Mv}{{\bf{M}}}
\newcommand{\Jv}{{\bf{J}}}
\newcommand{\Xv}{{\bf{X}}}
\newcommand{\xv}{{\bf{x}}}
\newcommand{\av}{{\bf{a}}}
\newcommand{\bv}{{\bf{b}}}
\newcommand{\cv}{{\bf{c}}}
\newcommand{\rv}{{\bf{r}}}
\newcommand{\nv}{{\bf{n}}}
\newcommand{\omegaell}{\omega_{\ell}}
\newcommand{\DOS}{\nu}
\newcommand{\vimp}{v_{\rm imp}}
\newcommand{\nimp}{n_{\rm imp}}
\newcommand{\nm}{n_{\rm m}}
\newcommand{\azero}{a_0}
\newcommand{\chiuni}{\chi_{0}}
\newcommand{\chiv}{\bm{C}}
%%%%%%%%%%%%%%%%%%%
\title{
Chirality driven anomalous Hall effect in weak coupling regime
} 
\author{Gen Tatara}
\author{Hikaru Kawamura}
\affiliation{
Graduate School of Science, Osaka University, Toyonaka, Osaka 560-0043, 
Japan
}
\date{\today}
%%%%%%%%%%%%%%%
\begin{abstract}
Anomalous Hall effect arising from non-trivial spin configuration 
(chirality) is studied based on the $s$-$d$ model.
Considering a weak coupling case, the interaction is treated perturbatively.
Scattering by normal impurities is included. 
Chirality is shown to drive locally Hall current and leads to overall
Hall effect if there is a finite uniform chirality.
This contribution is independent of the conventional 
spin-orbit contribution and shows distinct low temperature
behavior.
In mesoscopic spin glasses, 
chirality-induced anomalous Hall effect is  
expected below the spin-glass transition temperature.
Measurement of Hall coefficient would be useful 
in experimentally confirming the chirality ordering. 
\end{abstract}
\pacs{72.15.-v,75.30.-m,Ê75.50.Lk}
%72.15.-vÊÊÊ Electronic conduction in metals and alloys 
%75.30.-mÊÊ Intrinsic properties of magnetically ordered materials 
%75.50.LkÊÊ Spin glasses and other random magnetsÊ 
\maketitle

Hall effect in ferromagnetic metals has long been known to have 
anomalous component which does not vanish at zero external magnetic 
field.
Theories have explained this as due to the spontaneous magnetization 
and spin-orbit interaction\cite{Karplus54,Smit55,Luttinger58}. 
It was also shown based on the $s$-$d$ model that the high temperature 
(close to the critical temperature) behavior of the anomalous Hall 
effect is understood in terms of the fluctuation of the 
magnetization coupled with spin-orbit 
interaction\cite{Kondo62,Maranzana67}.

Recently some manganites were found to exhibit at high temperatures
abnormal behavior\cite{Snyder96,Asamitsu98} 
which is not explainable by previous theories. 
This behavior was explained by Berry phase effect 
associated with thermally driven non-trivial 
background spin configuration (chirality)\cite{Ye99,Chun00,Geller01}. 
Finite chirality results in a finite Berry phase and leads to 
Hall effect if chirality is non-vanishing as a net.
In contrast to manganites, anomalous Hall coefficient in 
ferromagnetic pyrochlores was found to remain finite at low 
temperatures\cite{Taguchi99,Katsufuji00}.
This behavior was discussed to be due to finite chirality of the 
ground state, which is originating from geometrical 
frustration\cite{Ohgushi00,Taguchi01,Shindou01,Onoda01}.
Behavior which is not explained solely by chirality 
theories was reported recently in some Mo-based 
pyrochlores\cite{Yoshii00,Yasui01,Kageyama01}.
These recent 
theories\cite{Ye99,Chun00,Ohgushi00,Taguchi01,Geller01,Shindou01,Onoda01} 
have exclusively 
dealt with a strong Hund-coupling limit, considering a half-metallic 
nature of the experimental systems.
In this limit, the electron spin aligns perfectly to the local 
spin and feels the same Berry phase as the local spin carries, and 
Hall conductivity has a topological meaning\cite{Thouless82}.
The weak coupling region, which would be the case of most common 
transition-metal magnets, has never been explored from 
the viewpoint of chirality.

In this paper, we study anomalous Hall effect due to chirality based 
on the $s$-$d$ model in the weak coupling case. 
Using Kubo formula and taking account of the impurity scattering, 
we will demonstrate that the chirality drives 
local Hall current in the perturbative regime.  
To compare with experiments, both the chirality mechanism and the 
conventional one 
due to the spin-orbit interaction (corresponding to the 
result by Karplus and Luttinger\cite{Karplus54,Luttinger58}) need to 
be taken account. 
We will show that the chirality contribution is independent of the 
conventional one, and they simply add up at lowest approximation.
In order for the chirality contribution to Hall effect to be finite, 
there needs to be a net uniform component of the chirality.
The possible effect of the spin-orbit interaction 
to induce a uniform component of the chirality in the presence of 
uniform magnetization, originally 
claimed to be present in the strong Hund-coupling limit\cite{Ye99}, 
is examined in our weak-coupling scheme. 
It turns out that the spin-orbit interaction induces a vector 
chirality if there is a uniform magnetization, and that this indeed  
results in a Hall effect in the bulk. 
This effect, however, would depend much on the band structure.

Our theory is applicable to a wide class of magnetic systems including 
canonical spin glasses, in which the 
conduction electron is only weakly coupled to the local spin.
In spin glasses, chirality order develops at low temperature leading 
to the spin-glass transition\cite{Kawamura92}.
Meanwhile, the chirality 
order there is spatially random  without a uniform component, making 
the experimental detection of the chirality-driven anomalous Hall 
effect rather difficult due to the inherent cancellation effect.
Even in this case, however, 
average of the squared chirality remains finite in small (mesoscopic)
samples\cite{Weissman93,Neuttiens00}, and 
enhancement of the anomalous Hall coefficient is expected below 
the spin-glass transition.
Thus, measurement of fluctuation of Hall conductivity in mesoscopic 
samples may be useful 
in experimental confirmation of the chirality ordering. 
Depending on the band structure, the chirality-driven Hall effect 
might be observable even in bulk spin-glass samples if the sample possesses a 
uniform magnetization, which is induced by applied fields or is 
generated spontaneously (as in case of reentrant spin 
glasses\cite{Mydosh93}).

%%%%%%%%%%%%%%%%%%%%%%%%%%%%%%%%%%%%%%%%%%%%%%%%%%%%%%%%%%%%%%%%%%%%%
We consider electron on lattice whose Hamiltonian is given by 
\begin{equation}
    H=\sum_{\kv\sigma}\epsilon_{\kv\sigma}c^\dagger_{\kv\sigma}c_{\kv\sigma}
    +\Hp+H_{\rm imp},
    \label{eq:Hdef}
\end{equation}
where $\sigma=\pm$ denotes electron spin. 
Electron energy is 
$\epsilon_{\kv\sigma}=\kv^2/2m-\sigma\Delta-\epsilon_{F}$, where 
$m$ is electron  mass, 
$\Delta$ is a uniform polarization of conduction electron 
 due to magnetization or external field, and
$\epsilon_{F}$ is the Fermi energy. 
The interaction with localized spin $\Sv_\Xv$ (treated as classical) is represented 
by the exchange interaction $\Hp$,
\begin{equation}
    \Hp=\frac{J}{N}\sum_{\kv\kv'} \Sv_{\kv'-\kv}  (c^\dagger_{\kv'}\sigmav c_{\kv}),
    \label{eq:Hp}
\end{equation}
where $\sigma^\alpha$ ($\alpha=x,y,z$) are Pauli matrices, and $N$ is 
the total number of lattice sites.
The sign of the exchange coupling $J$ depends on the system we consider; it is 
positive 
if the interaction is the $s$-$d$ exchange as in case of canonical 
spin glasses, and is negative if it is the Hund-coupling as in case of 
manganites.
Configuration of $\Sv_{\Xv}=(1/N)\sum_{\qv}e^{i\qv \Xv} \Sv_{\qv}$ is 
fixed. 
We note that localized spins $\Sv_{\Xv}$ do not necessarily occupy all $N$ 
sites here: An example is a dilute magnetic alloy such as canonical spin 
glass\cite{Mydosh93}.
$\Hp$ is assumed not to contain uniform component $\Mv\equiv \Sv_{q=0}$, 
since it is taken into account in $\Delta=JM$ with $M=|\Mv|$. 
The scattering by normal impurities is represented by 
$H_{\rm imp}=({\vimp}/{N})\sum \sum_{i}^{\nimp} e^{i(\kv-\kv')X_{i}}
c^\dagger_{\kv'}c_{\kv}$, where $\nimp$ is the number of nonmagnetic 
impurities (at sites $X_{i}$) and $\vimp$ is a constant.

The electronic current is given by 
$\Jv=\frac{e}{m}\sum_{\kv\sigma}\kv c^\dagger_{\kv'\sigma} c_{\kv\sigma}$.
Based on Kubo formula, the anomalous Hall conductivity is obtained as
$
\sigma_{xy}=\lim_{\omega\rightarrow0}\frac{1}{\omega}{\rm Im}
\left( Q_{xy}(\omega+i0)-Q_{xy}(i0) \right)
$
with 
$    Q_{xy}(i\omegaell)\equiv\frac{1}{\beta V} 
    <J_{x}(i\omegaell)J_{y}(-i\omegaell)>
$, 
where the bracket $<>$ denotes averaging over electrons and 
impurities, 
$\omega_{\ell}\equiv 2\pi\ell/\beta$ being the Matsubara frequency.
We treat $\Hp$ perturbatively.

\begin{figure}[bthp]
\includegraphics{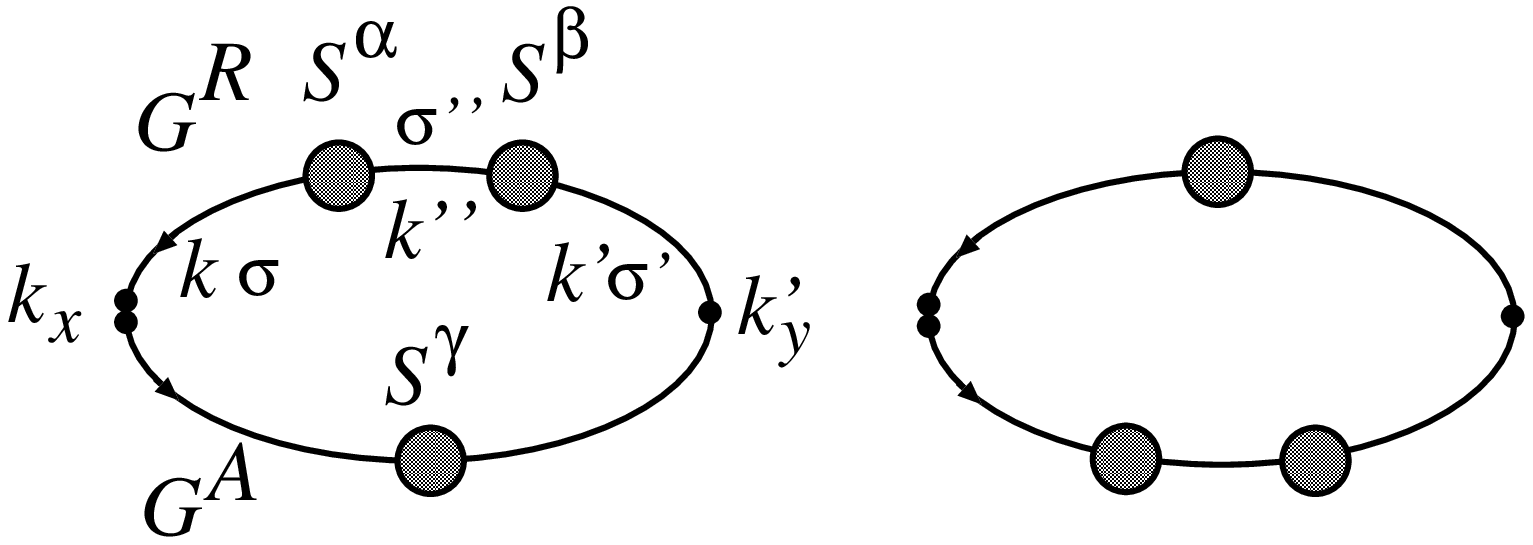} 
\begin{minipage}[t]{70mm}
\caption{Three spin contribution to $\sigma_{xy}$. The interaction 
with the local spin, $\Sv$, is denoted by a shaded small circles. 
The two processes are complex congugate to each other. 
Other contributions vanish due to symmetry. 
\label{FIGsigxy}}
\end{minipage}
\end{figure}

As is obvious, the first and second order contribution vanish since the spatial  
asymmetry due to
current vertices $J_{x}$ and $J_{y}$ in the correlation function cannot be 
deleted. 
The first term which can possibly be finite is the third-order term 
(Fig. \ref{FIGsigxy});
\begin{eqnarray}
  \sigma_{xy}^{(3)} &=& \frac{1}{2\pi V} 
  \left(\frac{e}{m}\right)^2 \left(\frac{J}{N}\right)^3 
  \sum_{\kv\kv'\kv''} 
  \sum_{\alpha\beta\gamma} k_{x} k'_{y} 
  S^{\alpha}_{\kv''-\kv} S^{\beta}_{\kv'-\kv''} S^{\gamma}_{\kv-\kv'}  
  {\rm tr}[
  G_{\kv}^R \sigma^\alpha G_{\kv''}^R \sigma^\beta
  G_{\kv'}^R G_{\kv'}^A \sigma^\gamma G_{\kv}^A ]
   \nonumber\\  &&   
   +{\rm c.c.},
  \label{eq:Q03def}
\end{eqnarray}
where $V$ is the total volume, trace is over spin indices, 
and $\alpha,\beta,\gamma$ runs over $x,y,z$.
$G_{\kv\sigma}^R(\equiv [{\frac{i}{2\tau}-\epsilon_{\kv\sigma}}]^{-1})$ and 
$G_{\kv\sigma}^A(=(G_{\kv\sigma}^R)^*)$ are retarded and advanced Green 
functions in the $\omega\rightarrow0$ limit, 
respectively, which include lifetime due to impurities, $\tau\equiv 
2\pi\DOS\nimp\vimp^2$, $\DOS$ being the density of states per site. 
We consider the case where the polarization of conduction electron is small 
($\Delta\sim 0$).
The summation over spin indices in eq. (\ref{eq:Q03def}) is then  
carried out as 
${\rm tr}[\sigma^\alpha \sigma^\beta \sigma^\gamma]= 
2i\epsilon^{\alpha\beta\gamma}$, where $\epsilon^{\alpha\beta\gamma}$ 
is the totally antisymmetric tensor. 
By use of the partial derivative, the Hall conductivity reduces to a 
compact form;
\begin{equation}
  \sigma_{xy}^{(3)}   = 
  \frac{N}{\pi V}
  \left(\frac{e}{m}\right)^2 (2\pi\DOS J)^3\tau^2 \chiuni
   =  (4\pi)^2 \sigma_{0} J^3 \DOS^2 \tau {\chiuni}, 
    \label{eq:Q0331}
\end{equation}
where $\sigma_{0}$ is the Boltzmann conductivity, 
$\sigma_{0}\equiv \frac{N}{2V}\left(\frac{e}{m}\right)^2\DOS 
k_{F}^2\tau$ ($k_{F}$ is the Fermi wavenumber).
We see that $\sigma_{xy}^{(3)}\propto \tau^2 \propto 
\rho_{0}^{-2}$ ($\rho_{0}=\sigma_{0}^{-1}$ is the resistivity).
The uniform chirality $\chiuni$ is given by 
\begin{eqnarray}
  \chiuni &\equiv&  \frac{1}{N}\sum_{\Xv_{i}}
    \Sv_{\Xv_{1}}\cdot(\Sv_{\Xv_{2}}\times\Sv_{\Xv_{3}})  
    \nonumber  \\
     &  & \times 
     \left[ 
     \frac{(\av\times\bv)_{z}}{ab} I'(a) I'(b) I(c) 
     +\frac{(\bv\times\cv)_{z}}{bc} I(a) I'(b) I'(c)
     +\frac{(\cv\times\av)_{z}}{ca} I'(a) I(b) I'(c)
     \right] , \label{chiunidef}
\end{eqnarray}
where $\Xv_{i}$ runs over all the positions of local spins, 
while $\av\equiv\Xv_{1}-\Xv_{2}$, 
$\bv\equiv\Xv_{2}-\Xv_{3}$ and $\cv\equiv \Xv_{3}-\Xv_{1}$ are the 
vectors representing sides of the triangle 
($a\equiv |\av|$ e.t.c.).
$I(r)\equiv 
\frac{1}{2\pi N\DOS\tau} \sum_{\kv}e^{i\kv\cdot\rv} G_{\kv}^R 
G_{\kv}^A$ and 
$I'(r)=\frac{d I(r)}{dr}$.
It is seen that Hall current is driven by three spins which form 
a finite solid angle in spin space (i.e., finite local chirality 
$\chi_{123}\equiv
\Sv_{\Xv_{1}} \cdot ( \Sv_{\Xv_{2}} \times \Sv_{\Xv_{3}}) $)
spanning a finite area in 
coordinate space (as seen from $(\av\times\bv)_{z}$ etc.).
Note that the factor in the square bracket in the definition of 
$\chiuni$ specifies the coupling between the spin- and 
the coordinate-space. 
In the case three spins ($1,2$ and $3$) align right-handed in 
spin space 
($\chi_{123}>0$),  $\chi_{123}$ contributes positively to $\chiuni$ 
if these three spins are 
located anti-clockwise in real space, and is negative if they are 
located clockwise.
In the case three spins align left-handed in spin space ($\chi_{123}<0$), this 
assignment is reversed.
Noting $I(r)=\frac{\sin k_{F}r}{k_{F} r}e^{-r/2\ell}$, 
where $\ell$ is elastic mean free path,
contribution from largely separated three spins
with the scale of $r$ decays rapidly as 
$\sim e^{-3r/2\ell}/(k_{F}r)^3$, and the Hall 
effect is dominantly driven by chiralities of spins on small triangles.
Note that the large-$r$ behavior of the weight function has 
resemblance to the RKKY interaction.
The expression of the uniform chirality 
derived in our weak coupling scheme, eq. (\ref{chiunidef}), 
contains contribution from large triangles, and is a natural 
extension of the conventional (and naive) definition of the 
chirality in terms of spins on adjacent sites only. 
Eqs. (\ref{eq:Q0331}) and (\ref{chiunidef}) are main results of 
the present paper, which gives a direct relation between the Hall 
conductivity and the spin configuration.

Conventional theories of anomalous Hall effect is based on the 
spin-orbit interaction, 
$H_{\rm so}\equiv
i\lambda\sum_{\kv\kv'}(\kv'\times\kv)\cdot 
    (c^\dagger_{\kv'}\sigmav c_{\kv})
$, where $\lambda$ is the spin-orbit coupling 
constant\cite{Karplus54,Smit55,Luttinger58}. 
Thus we have also analyzed the contribution of $H_{\rm so}$ on the same 
footing as that of $\Hp$ performing a double power series expansion.
At the lowest (first) order, the contribution of $H_{\rm so}$ 
is made up of so-called skew scattering and side-jump
ones\cite{Berger70}, which are calculated as 
$\sigma_{xy}^{\rm so}=-\lambda M (A'\tau+B)$\cite{Dugaev01}. 
Here $A'$ and $B$ are constants independent of $\tau$, each term 
corresponding to skew scattering and side-jump processes, respectively.
We note that $A'$ and $B$ are positive in the present single band 
approximation, but their signs actually 
depend on the band structure in real materials. 
At the lowest order, spin-orbit ($\sigma_{xy}^{\rm so}$) and chirality 
(eq. (\ref{eq:Q0331})) contributions are 
independent, and the total Hall conductivity is simply their sum.
These can mix as higher order corrections, but we neglect such small 
contributions.
The total Hall resistivity,  
$\rho_{xy}=\sigma_{xy}\rho_0^2$, 
then behaves as
\begin{equation}
\rho_{xy}\simeq -\lambda M(A \rho_{0}+B \rho_{0}^2) +C J^3 \chiuni,
\label{rhoxy}
\end{equation}
where $A=A'\rho_{0}\tau$ is independent of $\tau$, while
$C=\frac{1}{e^2}\frac{2V}{N}(\frac{m}{k_{F}})^2>0$ in our single band 
approximation. 
The sign of chirality contribution depends 
on whether the coupling is of the $s$-$d$ type ($J>0$) or of 
the Hund type ($J<0$). 
It is seen that the three terms in eq. (\ref{rhoxy}) 
depend differently on the impurity 
concentration. The chirality contribution is dominant in the 
clean regime and at low temperatures.
It should be noted that the analysis in the strong Hund-coupling 
case which does not consider impurities yields  
$\rho_{xy}\propto \rho_{0}^2 \chi$\cite{Ohgushi00}.
These different dependences on $\rho_{0}$, which indicate different 
behavior as a function of temperature, would be useful in interpreting 
the experimental results.

The chirality contribution to Hall coefficient is finite only if 
there is a net uniform chirality, $\chiuni\neq0$.
Finite net chirality, however, may not be very easy to realize on regular 
lattices with simple nearest-neighbor exchange interaction, 
since the chirality on adjacent plaquettes usually tends to cancel each 
other due to symmetry\cite{Ohgushi00,Shindou02}.
One possible mechanism to realize a finite net chirality has been 
proposed in Ref. \cite{Ye99}, where it was argued in the strong 
coupling case that 
the spin-orbit interaction induced a net chirality in the 
presence of magnetization as
$\chiuni^{\rm so}=-\alpha\lambda M$ ($\alpha$ is a positive 
constant).
Inspired by this observation, 
we have examined whether such a mechanism works in the present weak 
coupling case.  
To examine the possible coupling between the uniform chirality and the 
magnetization $M$, we look into the expectation value (effective 
Hamiltonian) of the spin-orbit 
interaction, $<H_{\rm so}>$, where $<>$ denotes the thermal averaging over 
electrons,  treating $\Hp$ as perturbation.
We identify two types of terms as possible candidates. One is the 
term linear in $M$ for small $M$ and comes from the third-order 
contribution in $\Hp$,
\begin{eqnarray}
    H_{\rm so}^{(3)\chi} &=&  2\lambda\left(\frac{J}{N}\right)^3 
     \sum_{\xv\Xv_{i}} 
    \Sv_{\Xv_{1}} \cdot ( \Sv_{\Xv_{2}} \times \Sv_{\Xv_{3}}) 
    [(\Xv_{1}-\xv)\times (\Xv_{3}-\xv)]_{z} 
    \nonumber\\
    && \times \frac{1}{\beta}\sum_{\omega_{n}}
     g_{\omega_{n}}'(|\xv-\Xv_{1}|)
 g_{\omega_{n}}(a)  g_{\omega_{n}}(b) \Delta'_{\omega_{n}}(|\Xv_{3}-\xv|),
     \label{eq:Hsoeff2}
\end{eqnarray}
where $\xv$ denotes the site at which the spin-orbit interaction acts 
($a$ and $b$ are defined after eq. (\ref{chiunidef})), and the uniform 
magnetization is assumed to be in $z$-direction.
The thermal Green functions are defined here as  
$g_{\omega_{n}}(r)\equiv \frac{1}{2}(G_{\omega_{n} +}+G_{\omega_{n} -})$, 
$\Delta_{\omega_{n}}(r)\equiv (G_{\omega_{n} +}-G_{\omega_{n}-})$, where 
$G_{\omega_{n}\sigma}(\Xv)=\sum_{\kv}{e^{-i\kv \Xv}} 
[{i(\omega_{n}+\frac{{\rm 
sgn}(\omega_{n})}{2\tau})-\epsilon_{\kv}+\Delta\sigma}]^{-1}$, and 
$G'(r)\equiv\frac{dG}{dr}$.
Note that 
$\Delta_{\omega_{n}}$ is proportional to $\Delta$, and hence to $M$,  for 
$\Delta/\epsilon_{F}\ll1$.
It is seen that 
$H_{\rm so}^{(3)\chi}$ apparently describes the coupling between the uniform 
magnetization and the local 
chirality, but without referring to the spatial spin configuration.
In fact, since the factor of $((\Xv_{1}-\xv)\times (\Xv_{3}-\xv))_{z}$ 
specifies only the angles between 
the two spins $S_{\Xv_{1}}$ and $S_{\Xv_{3}}$ when looked from the  
position $\xv$ irrespective of spatial configuration 
of $S_{\Xv_{2}}$, $H_{\rm so}^{(3)\chi}$ does not contain the 
component inducing the uniform chirality $\chiuni$.

The other term is quadratic in $M$ for small $M$ and comes from the 
second-order contribution in $\Hp$, which reads as
\begin{eqnarray}
    H_{\rm so}^{(2)\chi} &=& 
    -2\lambda\left(\frac{J}{N}\right)^2 
    \sum_{\xv\Xv_{i}} (\Sv_{\Xv_{1}}\times \Sv_{\Xv_{2}})_{z} 
    \frac{[(\Xv_{1}-\xv)\times(\Xv_{2}-\xv)]_{z}}
       {|\Xv_{1}-\xv| |\Xv_{2}-\xv|}  \nonumber\\
       && \times
   \frac{1}{\beta}\sum_{\omega_{n}} 
    \Delta_{\omega_{n}}'(|\Xv_{1}-\xv|) g_{\omega_{n}}(|\Xv_{1}-\Xv_{2}|) 
    \Delta_{\omega_{n}}'(|\Xv_{2}-\xv|).
    \label{hso2}
\end{eqnarray}
In the presence of uniform magnetization $M$, 
$z$-component of the vector chirality, defined 
by two spins as $\chiv_{12}\equiv(\Sv_{\Xv_{1}}\times\Sv_{\Xv_{2}})$, 
plays essentially the same role as the scalar chirality 
($\chi_{123}\simeq M\chiv_{12}^{z}$).   
Then, eq. (\ref{hso2}) above describes the coupling between the vector 
chirality and the magnetization via the spin-orbit interaction.
After summation over the thermal frequency, 
the Fourier transform of the electron part becomes
\begin{equation}
    \frac{1}{\beta}\sum_{\omega_{n}} 
    \Delta_{\omega_{n}\kv}g_{\omega_{n}\kv'}\Delta_{\omega_{n}\kv''}=
    \frac{1}{8}\sum_{\sigma\sigma'\sigma''}
    \frac{\sigma\sigma''}{\epsilon_{\kv\sigma}-\epsilon_{\kv'\sigma'}}
    \left[ 
 \frac{f_{\kv\sigma}-f_{\kv''\sigma''}}
 {\epsilon_{\kv\sigma}-\epsilon_{\kv''\sigma''}}
- \frac{f_{\kv'\sigma'}-f_{\kv''\sigma''}}
 {\epsilon_{\kv'\sigma'}-\epsilon_{\kv''\sigma''}}
\right],
 \label{eq:sumzero}
\end{equation}
where $f_{\kv\sigma}$ is the Fermi distribution function.
If there is a particle-hole symmetry (invariance under
$\epsilon\rightarrow-\epsilon$), which is equivalent to assuming that 
only excitations close to the Fermi level dominates, 
eq. (\ref{eq:sumzero}) after summation over $\kv$'s turns out to vanish.
This symmetry, however, is not necessarily observed in real material,  
for instance due to 
the existence of the bottom of the band, and then, a finite contribution 
is expected to remain in general. 
Thus, this term could serve as a ``symmetry-breaking field'' 
($H_{\rm so}^{(2)\chi}\propto M^2\sum_{\Xv_{i}}\chiv_{12}^{z} 
\propto M \chiuni$) inducing a uniform chirality, which results 
in a Hall effect in the bulk. 

%%%%%%%%%%%%%%%%%%%%%%%%%
Even in the case the chirality does not contain uniform component,  
$\chiuni$ could still 
be finite if the system size is sufficiently small. 
Let us consider the case of spin glasses in zero external field, 
in which the chirality is randomly ordered. 
The number of triangles which contribute to $\sigma_{xy}$ 
is given roughly as 
$N_{\chi}\simeq N \nm^3 (\ell/\azero)^4$  ($\nm$ is the concentration 
of localized spins, and $\azero$ is lattice constant). 
Hence, the sum of random chirality is
$\chiuni\propto\frac{1}{\sqrt{N}}\nm^{3/2} (\ell/\azero)^2$. 
Although this quantity decays as $\propto 1/\sqrt{N}$, 
we expect that detection of Hall effect 
may be possible by high-sensitivity-measurements on mesoscopic samples. 
This chirality-driven Hall effect of random sign in mesoscopic spin glasses
would be measurable\cite{Weissman93,Neuttiens00}
by looking at the sample-dependent or 
 thermal-cycle-dependent fluctuations, 
$\delta\sigma_{xy}\equiv \sigma_{xy}-[{\sigma_{xy}}]_{s}$ 
($[\;]_{s}$ denotes average over spin configurations), 
whose squared average is given as
$\sqrt{[{ (\delta \sigma_{xy})^2}]_{s}} /\sigma_{0}
 = (4\pi)^2 J^3 \DOS^2 \tau
  \sqrt{ [{\chi^2}]_{s} } $, 
where
  $\sqrt{[{\chi^2}]_{s}}\equiv 
{ \frac{1}{N} \{ \sum_{ijk} 
(\chi_{ijk} F_{ijk})^2 }\}^{1/2}
\propto\frac{1}{\sqrt{N}}\nm^3 (\ell/\azero)^4$ 
($F_{ijk}$ represents the spatial weight given in the square bracket 
in eq. (\ref{chiunidef})).

%%%%%%%%%%%%%

In the 70's, anomalous Hall effect in spin glasses was experimentally 
investigated\cite{Mydosh93,McAlister76}. 
It was found that the Hall resistivity 
of canonical spin glasses, i.e., dilute magnetic alloys 
such as AuFe and AgMn, measured in weak
applied fields were negative and exhibited a cusp-like anomaly 
around the spin-glass transition 
temperature whose behavior was quite similar to that of the 
magnetic susceptibility, $\chi_{m}$; i.e., 
$\rho_{xy}/H\propto-\chi_{m}$\cite{McAlister76}.
Although this behavior can be explained by  
the standard spin-orbit contribution  (the first two terms in eq. 
(\ref{rhoxy})),  it could also be explainable 
by the contribution of a uniform chirality 
induced by $H_{\rm so}^{(2)\chi}$ described above.
In order to experimentally resolve the spin-orbit contribution of
Karplus-Luttinger type  from the spin-orbit induced chirality 
contribution, one might possibly examine the dependence on 
$\rho_{0}$ (Eq. (\ref{rhoxy})), or measure  
the response to external magnetic fields applied  
in various directions.
In this connection, reentrant spin-glass systems\cite{Mydosh93}, which exhibit 
successive phase transitions, first from para to ferro 
and then from ferro to spin glass at lower temperature, would be of 
much interest. In the ferromagnetic regime, only 
the conventional spin-orbit mechanism is expected to work, while in 
reentrant spin-glass regime, the chirality contribution 
sets in due to the spin canting giving rise to distinct contribution 
at lower temperatures.

%%%%%%%%%%%%%%%%%%%%%%%%%%

To summarize, we have demonstrated based on the $s$-$d$ model in the weak 
coupling 
regime that a topologically nontrivial spin configuration (chirality) induces 
Hall current.
The chirality-driven Hall effect as a bulk appears if uniform component 
of the chirality is finite.
This contribution is independent of the conventional 
contribution from the spin-orbit 
coupling exhibiting different dependence on resistivity from the 
spin-orbit contribution and other predictions based on chirality 
mechanism.
If chirality is ordered randomly, as in spin glasses below the spin-glass 
transition temperature,
sample-to-sample or thermal-cycle-dependent fluctuations of Hall conductivity
in mesoscopic samples is expected to show an anomalous enhancement.
Without a direct method of magnetic detection of the chirality 
available so far, 
measurement of (fluctuation of) Hall conductivity would be powerful tool  
in experimental confirmation of the chirality ordering. 

%%%%%%%%%%%%%%%%%%%
%\acknowledgements
\begin{acknowledgements}
The authors are grateful to H. Kohno, R. Shindou, H. Yoshino, S. 
Yoshii, N. Nagaosa and M. Sato
for useful comments and discussion.
This work is supported by Priority Areas Grants from the 
Ministry of Education, Culture, Sports, Science and Technology, Japan.
G. T. thanks The Mitsubishi Foundation for financial support.
\end{acknowledgements}

%%%%%%%%%%%%%%%%%%%%

%%%%%%%%%%%%%%%%%%%%%%%%%%%%%%%%%%%%%%%%%%%%%%%%%%%%%%%%%%%%%%%%%%%%%
\end{document}